\documentclass[epj]{svjour}
\usepackage{graphics}

\usepackage[english]{babel}
\usepackage{amsmath,amssymb}
\usepackage{times}

\begin{document}
\title{Violation of the Luttinger sum rule within the Hubbard model 
on a triangular lattice}

\author{J. Kokalj\inst{1} and P. Prelov\v sek\inst{1,2}}
\institute{J.\ Stefan Institute, SI-1000 Ljubljana, Slovenia \and
 Faculty of Mathematics and Physics, University of
 Ljubljana, SI-1000 Ljubljana, Slovenia}

\date{Received: \today / Revised version: date}
                   
\abstract{
The frequency-moment expansion method is developed to analyze the
validity of the Luttinger sum rule within the Mott-Hubbard insulator, 
as represented by the generalized Hubbard model at half filling and
large $U$. For the particular case of the Hubbard model with
nearest-neighbor hopping on a triangular lattice lacking the
particle-hole symmetry results reveal substantial violation of the sum
rule.
}
\maketitle

\section{Introduction}
\label{sec:intro}
The concept of a Fermi liquid (FL) is the basic one for the
understanding of electrons in solid state. The central pillar of the
FL theory is the existence of a Fermi surface with a volume unchanged
by electron-electron interactions, having the firm support in the
Luttinger sum rule (LSR) \cite{luttinger60b,luttinger60a,abrikosov}.
In last two decades experiments on novel materials with strongly
correlated electrons, in particular superconducting cuprates
\cite{yoshida06}, and theoretical analyzes of corresponding microscopic
electronic models revived the question of the possible limitations and
the breakdown of the LSR.

The LSR within a paramagnetic metal relates the density of electrons
to the Fermi surface (${\bf k}$ space) volume defined by poles of the
Green's function (GF) $G({\bf k},\omega=0)$. Only recently it has been
recognized that the concept can be generalized to insulators
\cite{dzyalosh03,konik06} whereby the corresponding 'Luttinger'
surface is defined by the zeros of $G({\bf k},0)=0$. It has been shown
that the LSR is indeed satisfied for models with a particle-hole
symmetry at half filling, and in particular for the frequently invoked
Hubbard model on a one-dimensional (1D) chain and on a 2D square
lattice \cite{stanescu07}. There are several indications that in the
absence of particle-hole symmetry the LSR can be generally violated
\cite{stanescu07,rosch07} although no explicit example emerging from a
microscopic model with a repulsive Coulomb interaction term
and within the canonical ensemble (assuming fixed particle density)  
has been presented so far. On the other hand, such cases seem to be
found for a Hubbard model on an inhomogeneous lattice and on a
homogeneous lattice with odd number of sites close to half filling
\cite{ortloff07}. Recently, the present authors
pointed out on the generalization of the LSR to finite systems of
interacting electrons \cite{kokalj07}. Since the original proof
\cite{luttinger60a,luttinger60b} of the LSR remain valid even for
small systems, this allows an alternative way to show and understand a
possible breakdown of LSR. So far, results confirm cases of violation
in restricted (nonperturbative) models as the $t$-$J$ model, whereas 
reachable systems and discussed model regimes did not provide a
clearcut violation within the translationally symmetric Hubbard model
\cite{kokalj07}. 

In this paper we analyze in more detail the validity of the LSR within
the insulating state emerging from a single-band metal via the
Mott-Hubbard mechanism of strong electron-ele\-ctron repulsion.
As the prototype model we use the Hubbard model on triangular lattice
with the large 
onsite repulsion $U \gg t$. We develop an approach employing the
expansion of frequency moments in powers of $t/U$ to determine the
value of the $G({\bf k},0)$ leading to the location of the
Luttinger surface $G({\bf k}_L, 0)=0$. The method is valid in the
thermodynamic limit (for an infinite system) but can be tested as well
on small systems, which can provide also required static quantities as
input. While the method confirms the validity of LSR in Hubbard models
with a particle-hole symmetry \cite{stanescu07}, we further
concentrate on the Hubbard model on a triangular lattice as the case
without the latter symmetry. Results show that the LSR is violated
substantially for large $U/t$, being an indication for analogous
violation in more general Mott-Hubbard insulators without
particle-hole symmetry.

For simplicity we restrict in the following our discussion to the
single-band Hubbard model on a general lattice
\begin{equation}
H = - \sum_{ijs} t_{ij} c_{js}^\dagger c_{is}+ U \sum_i n_{i
\uparrow} n_{i\downarrow}, \label{eq6}
\end{equation}
where $U$ is the local Coulomb repulsion, while as the intersite
hopping $t_{ij}$ we can consider besides the nearest neighbor (n.n.)
term $t_{ij}=t$ also extended model, e.g., with the next-nearest
neighbor (n.n.n.) hopping $t_{ij}=t'$. We are interested in the regime
of large enough $U \gg t_{ij}$ and half filling $n=1$, where the
ground state is insulating with a Mott-Hubbard gap separating both
Hubbard bands. Reduced models, such as the $t$-$J$ model, are mostly
adequate enough to discuss physics in this situation. The essential
motivation to stay with the Hubbard model is the fact that the latter
model is perturbative, at least for small $U$, and thus could be
adiabatically connected to the case of noninteracting ($U=0$)
fermions. The breakdown of the LSR could be then attributed to
nonadiabatic development of the ground state (i.e. in the limit
$T \to 0$) through the metal-insulator transition, allthough there
could be also other interpretations. 

In the following we first derive lowest orders of GF with the use of
large $U$ expansion method. We also use particle-hole symmetry and
give in some more detail the calculation of chemical potential. In
chapter \ref{sec:results} we first give general results of large $U$
expansion method and in chapter \ref{sec:hubbard} we present our
results for particular case of triangular lattice. At the end a
discussion of results and conclusions are given.

\section{Large $U$ expansion}
\label{sec:large}
The central quantity for the LSR is the GF at temperature $T=0$, given
by
\begin{eqnarray}
G({\bf k},\omega)&=&-i\int_{-\infty}^0 dt \mathrm{e}^{i(\omega+\mu)t}
\langle0| c_{{\bf k}s}^\dagger c_{{\bf k}s}(t) |0\rangle \nonumber\\
&&+i\int_0^\infty dt \mathrm{e}^{i(\omega+\mu)t} \langle 0| c_{{\bf
k}s}(t) c_{{\bf k}s}^\dagger |0\rangle,
\label{GWithC}
\end{eqnarray}
where $\mu$ is the chemical potential. Representing the GF in terms
of the spectral function
\begin{equation}
G(\mathbf{k},\omega)=\int_{-\infty}^{\infty}
\frac{A(\mathbf{k},\omega')}{\omega+\mu - \omega' \pm i \eta} d\omega'.
\end{equation}
We deal with the insulator with real $G(\mathbf{k},\omega \sim 0)$ and
a large gap $\Delta \sim U$ separating regions of nonzero spectral functions
$A(\mathbf{k},\omega)$. In order to evaluate the relevant $G({\bf k},0)$
we separate contributions
\begin{eqnarray}
&&G(\mathbf{k},0)=G^-(\mathbf{k},0)+G^+(\mathbf{k},0)= \nonumber \\
&=&\int_{-\infty}^{\mu}
\frac{A(\mathbf{k},\omega')}{\mu - \omega' - i \eta} d\omega'
+\int_{\mu}^{\infty}
\frac{A(\mathbf{k},\omega')}{\mu - \omega' + i \eta} d\omega'
\label{GintA}
\end{eqnarray}
Since $\mu \sim U/2$ it is reasonable to expand both
$G^-(\mathbf{k},0)$ and $G^+(\mathbf{k},0)$ in powers of
$(t_{ij}/U)^n$.

Let us first consider $G^-$,
\begin{eqnarray}
G^-(\mathbf{k},0)&=&\sum_{n=0}^\infty \bigl(\frac{2}{U} \bigr)^{n+1}
\int_{-\infty}^{\mu}
A(\mathbf{k},\tilde{\omega}')(\tilde \omega'-\tilde \mu)^n
d\tilde{\omega}' = \nonumber \\
&=&\sum_{n=0}^\infty \bigl(\frac{2}{U} 
\bigr)^{n+1}  \sum_{m=0}^n M^-_{n-m}(\mathbf{k}){n \choose m} (-\tilde \mu)^m,
\label{GMinus}
\end{eqnarray}
where $\tilde \mu = \mu-U/2$. Note that $M^-_l({\bf k})$ are frequency
moments which can be directly expressed in terms of the equal-time
expectation values within the half-filled ground state $|0\rangle $
involving $l$ commutators with $H$,
\begin{equation}
M^-_l({\bf k})=\langle[H,\ldots [H, 
c^\dagger_{{\bf k}s}]] c_{{\bf k}s} \rangle.
\end{equation}
Analogous procedure can be repeated for $G^+$ with the result
\begin{eqnarray}
G^+(\mathbf{k},0)&=&\sum_{n=0}^\infty \bigl(-\frac{2}{U} \bigr)^{n+1}
\sum_{m=0}^n M^+_{n-m}(\mathbf{k}){n \choose m} ( -\tilde \mu)^m, \nonumber \\
M^+_l({\bf k})&=&\langle c_{{\bf k}s} 
[H',\ldots [H',c^\dagger_{{\bf k}s}]] \rangle,
\label{GPlus}
\end{eqnarray}
where $H'=H-U(\hat N_e-N)$, with $\hat N_e$ being electron-number
operator and $N$ is the number of sites in the system. $G^+$ is also
tightly related to $G^-$ with the particle-hole transformation, as
discussed lateron.

Let us now present first two moments explicitly,
\begin{eqnarray}
M^-_0(\mathbf{k})&=&\langle c^\dagger_{\mathbf{k}s} 
c_{\mathbf{k}s}\rangle=\bar n_{{\bf k}s}, \nonumber \\
M^+_0(\mathbf{k})&=&\langle c_{\mathbf{k}s} 
c^\dagger _{\mathbf{k}s}\rangle=1-\bar n_{{\bf k}s}, \nonumber \\
M^-_1(\mathbf{k})&=&\varepsilon(\mathbf{k}) \bar n_{\mathbf{k}s} 
  + \frac{U}{N}\sum_{ij}e^{-i\mathbf{k}(i-j)}\langle
 c_{i s}^\dagger n_{i\bar{s}} c_{js} \rangle, \nonumber \\
M^+_1(\mathbf{k})&=&\varepsilon(\mathbf{k})(1-\bar n_{\mathbf{k}s}) +
\nonumber \\
&&
+ \frac{U}{N}\sum_{ij}e^{-i\mathbf{k}(i-j)}\langle
 c_{j s} (n_{i\bar{s}}-1) c^\dagger _{i s} \rangle,
\label{M01untr}
\end{eqnarray} 
where $\varepsilon(\mathbf{k})$ is the free band ($U=0$) dispersion as
determined by hopping parameters $t_{ij}$.

One expects that for large $U\gg t_{ij}$, $M^-_l$ is of the order of
$t_{ij}^l$, at least for lowest orders $l$,
as well as $\tilde \mu \propto t_{ij}$. Hence the series 
Eq.(\ref{GintA}) should be well converging. To show that $M^-_l
\propto t_{ij}^l$, we employ at this
stage the well known canonical transformation of the Hubbard model for
large $U$ \cite{harris67,macdonald88,eskes94}, with which the number of
doubly occupied sites becomes a good quantum number. This leads at
half-filling, $n=1$, to an effective 
spin model \cite{macdonald88} and at finite doping to strong coupling (SC)
model for the lowest orders \cite{eskes94}.
\begin{equation}
H_{\textrm ef}=e^S H e^{-S}, \qquad a_{is}=e^S c_{is} e^{-S}.
\label{extj}
\end{equation}
Within the lowest order of the $t_{ij}/U$ expansion one gets
$S=S^1$,
\begin{equation}
 S^1\!=\!-\frac{1}{U}\!\sum_{i j s} t_{ij} [
 n_{i \bar s} c_{i s}^\dagger c_{js} (1\!-\!n_{j\bar s})
 -\!(1\!-\!n_{i \bar s}) c_{is}^\dagger c_{js} n_{j \bar s}
 ].
\label{awithc}
\end{equation}

Transforming expressions for moments (\ref{M01untr}) and using
Eq. (\ref{awithc}), we arrive to explicit expressions at $n=1$, 
\begin{eqnarray}
M^\mp_0(\mathbf{k})= \frac{1}{2} \pm \!\frac{2}{U}\! \sum_{\delta}
\varepsilon_\delta(\mathbf{k}) [\langle \mathbf{S}_{\delta}\cdot
\mathbf{S}_0 \rangle \!-\!\frac{1}{4}]\! + \!O(\frac{t_{ij}^2}{U}),
\nonumber \\
M^\mp_1(\mathbf{k})= \frac{\varepsilon(\mathbf{ k})}{2}
+\sum_{\delta} \varepsilon_\delta(\mathbf{k}) [\langle
\mathbf{S}_{\delta}\cdot \mathbf{S}_0 \rangle\! -\!\frac{1}{4}]\! +
\!O(\frac{t_{ij}^2}{U}),
\label{MExpan}
\end{eqnarray}
where $\varepsilon_\delta({\bf k})$ refers to partial 'free' bands
corresponding to particular neighbors (n.n., n.n.n. etc.),
respectively
\begin{equation}
\varepsilon_\delta(\mathbf{k})= -t_{\delta} \sum_{i_\delta}
\mathrm{e}^{i\mathbf{k} \mathbf{r}_{i\delta}}.
\label{epsilonk}
\end{equation}
We note that spin correlations $\langle \mathbf{S}_{\delta}\cdot
\mathbf{S}_0 \rangle$ in Eq. (\ref{MExpan}) are to be evaluated in the
ground state of $H_{\textrm ef}$ and that such state does not contain doubly
occupied sites. 

\subsection{Particle-hole symmetry}
\label{sec:particle}
Let us now return to $G^+$. In order to show that its moments are
closely related to $M_l^-$, it is very instructive to realize that
$G^+$ can be obtained from $G^-$ with particle-hole transformation,
which changes $c_{is} \to c_{is}^\dagger$ and vice versa. Applying the
transformation to $G^+$ in Eq. (\ref{GWithC}) we note that also the
ground state $|0\rangle$ changes to $|\bar 0\rangle$, corresponding to
transformed $\bar H$.  $\bar H$ is obtained from $H$ by replacing
$t_{ij}\to -t_{ij}$ and adding the extra potential term $U(N-N_e)$. We
therefore arrive to the following expression for $G^+$,
\begin{equation}
G^+({\bf k},0)=i\int_{-\infty}^o dt \mathrm{e}^{i(U-\mu)t} \langle
\bar 0| c_{{\bf k}s}^\dagger \mathrm{e}^{i\bar H t} c_{{\bf k}s}
\mathrm{e}^{-i \bar H t}| \bar 0 \rangle.
\end{equation}
which is analogous to $G^-$ in Eq.(\ref{GWithC}). One can then
continue by analogy to $G^-$ using the unitary transformation with
$\bar S$, which is the same form as $S$ with $-t_{ij}$ instead of
$t_{ij}$. State, for which expectation value must be calculated, is
now the ground state of $\bar H_{\textrm ef}$. Moreover, for half filling at
$n=1$, $H_{\textrm ef}$ contains only spin operators, whereby odd spin terms
(also odd terms of $t_{ij}$) are forbidden by the isotropy of
original Hamiltonian, and $\bar H_{\textrm ef} = H_{\textrm ef}$ \cite{macdonald88},
and therefore also ground states of them are the same. Finally we can
relate
\begin{equation}
G^+({\bf k}, 0)= -\bar G^-({\bf k}, 0)_{\tilde \mu \to -\tilde \mu},
\label{GpwithGm}
\end{equation}
where $\bar G^-$ stands for $G^-$ with $t_{ij}$ changed to
$-t_{ij}$ and $\tilde \mu \to -\tilde \mu$. Eq. (\ref{GpwithGm}) 
confirms that $M_l^+$ are simply related to $M_l^-$ as already
seen from relations (\ref{MExpan}).

Expressions are so far valid for general hopping $t_{ij}$, with the
only assumptions that we are dealing with the paramagnetic state with
total spin $S=0$ and without broken translational symmetry.

\subsection{Chemical potential}
\label{sec:chemical}
We are interested in the case of Mott-Hubbard insulator at half
filling, n=1 and within the canonical ensemble the
chemical potential $\mu$ should be determined from the grand partition
function by first fixing the number of particles in the system and
then taking the limit $T\to 0$. Such a
procedure gives an unique value for $\mu$ in the middle of the
gap \cite{rosch07,perdew82,kaplan06}, and
works as well for finite systems \cite{kokalj07}.  Different limits
with $\mu$ being anywhere within the
Mott-Hubbard gap have been also recently considered
\cite{rosch07,ortloff07}. Our procedure simply locates $\mu$
(as well as in any finite system with $N_e$ particles \cite{kokalj07})
with the ground state energies of systems with one electron more and
less, respectively \cite{rosch07,perdew82,kaplan06},
\begin{equation}
\mu = (E_0^{N_e+1}-E_0^{N_e-1})/2
\end{equation}
where for
half-filling $N_e=N$. Since the ground state of one hole for 
$H_{\textrm ef}$ does not contain doubly occupied sites, we get
$E_0^{N-1}=O(t_{ij})$.  On the other hand, ground state with one added
electron more contains one doubly occupied site and therefore
$E_0^{N+1}=U+O(t_{ij})$. $\mu$ is therefore within the gap and
approximately $\mu \sim U/2$. Energies $E_0^{N+1}$, $E_0^{N-1}$ as
well as $\tilde \mu$ might be expanded analytically in terms of
$t_{ij}/U$, however we use furtheron numerical results instead.

\section{Results}
\label{sec:results}
Using expanded moments (\ref{MExpan}) in $G^-$ (\ref{GMinus}) and
Eq. (\ref{GpwithGm}) for $G^+$, we finally get for the total GF
(\ref{GintA}),
\begin{equation}
G(\mathbf{k},0)=
  \frac{4}{U^2}
  (\sum_\delta 4 \varepsilon_\delta(\mathbf{k})
   \langle \mathbf{S}_{\delta}\cdot \mathbf{S}_0 \rangle
   -\tilde \mu) + O(\frac{t_{ij}^2}{U^3}),
   \label{GExpan}
\end{equation}
From Eq. (\ref{GintA}) and (\ref{GpwithGm}) one can also conclude,
that for $\tilde \mu =0$, GF consists only of odd terms in $t_{ij}$.

It has been already realized \cite{stanescu07}, that the LSR remains
valid for cases with particle-hole symmetry, where $\bar H=H$.  This
follows also from our analysis. If we assume in Eq. (\ref{GExpan})
only nearest neighbors and put $\tilde \mu=0$, which is the
consequence of the particle-hole symmetry, GF changes sign at same
$\mathbf{k}$ as $\varepsilon(\mathbf{k})$. Within the lowest order in
$t_{ij}/U^2$, this leads to the same Luttinger volume as for
noninteracting fermions.

In order to find cases where the LSR might be violated, we 
turn to systems without the particle-hole symmetry. Here, in general
$\tilde \mu \ne 0$ and also $\varepsilon(\mathbf{k})$ is not negative
for exactly half of the Brillouin zone. Then, Eq. (\ref{GExpan}) seem to
suggest that the LSR must be violated in most cases, since the perfect
cancellation of two quite different terms in Eq. (\ref{GExpan}) looks quite
improbable. In the following, we show that such compensation indeed
does not occur for one of simplest non-symmetric cases, i.e. within
the Hubbard model with only n.n. hopping on the triangular lattice.

\subsection{Hubbard model on a triangular lattice}
\label{sec:hubbard}
Furtheron we
consider the model with n.n. hopping only, $t_{ij}=t$. To check the
LSR on triangular lattice at least to lowest order in $t/U$, some
quantities are needed. Free band dispersion $\varepsilon(\mathbf{k})$
is given with Eq. (\ref{epsilonk}) and equals
$\varepsilon_1(\mathbf{k})$.  Spin correlations $C_{\delta} =\langle
\mathbf{S}_{\delta}\cdot \mathbf{S}_0 \rangle$ should be evaluated in
the ground state of $H_{\textrm ef}$ at half-filling.  Within the
first order in $t^2/U$ this corresponds just to the Heisenberg model
on a triangular lattice.  The value for n.n. spin correlations in
Heisenberg model has been evaluated by a number of authors
\cite{bernu92,capriotti99,koretsune02} and $C_1 \doteq -0.182$. This
value is quite insensitive to finite $t^2/U$, e.g. we obtain
within the Hubbard model on $N=15$ sites $-C_1 \sim 0.183 - 0.189$ for
$t^2/U=0.05~t - 0.02~t$.

\begin{figure}
\resizebox{0.5\textwidth}{!}{%
\includegraphics {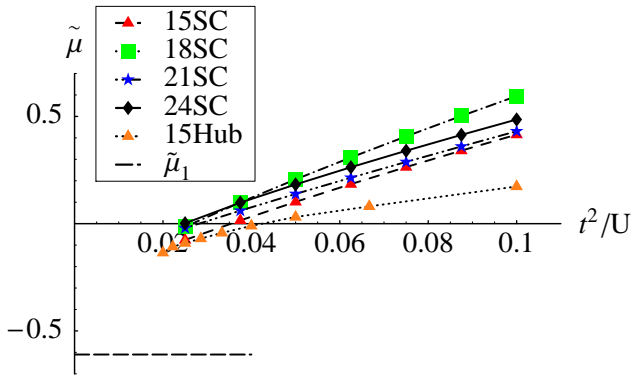}
}
\caption{(Color online) Chemical potential $\tilde \mu$ on a
triangular lattice vs. $t^2/U$ calculated on clusters with $N=15 - 24$
sites within the Hubbard model (15Hub) and in strong coupling (SC)
models, respectively. The value $\tilde \mu_1$ for which LSR would be
satisfied is also shown.}
\label{muplot}
\end{figure}

The chemical potential $\tilde \mu$ is more delicate quantity. Here,
we estimate $\tilde \mu$ as accurate as possible by calculating
$E_0^{N+1}$ and $E_0^{N-1}$, performing the exact diagonalization of
finite clusters using the Lanczos algorithm. Lattices with periodic
boundary conditions are chosen in such way, that they allow
three-sublattice order \cite{bernu92} and have a minimal imperfection
\cite{kent05}, that is zero for all choosen lattices, except for N=18
it is 2. The maximum size
for the Hubbard model is $N=15$, for which we present $\tilde \mu$ in
Fig. \ref{muplot}. Fortunately, for larger $U/t$ the Hubbard model can be well
represented by the SC model (equivalnet to the $t$-$J$ model
additionally including three-site
correlated hopping term of the order $t^2/U$) \cite{eskes94}.
As seen in Fig.~\ref{muplot}, $\tilde \mu$
calculated from the Hubbard model and from the SC model for $N=15$
match well for small $t^2/U<0.05$. This gives support to the values
obtained for $\tilde \mu$ within the SC model, as presented in
Fig. \ref{muplot} for systems up to $N=24$.

In considering $\tilde \mu$ one should take into account that at large
$U/t$ the Nagaoka ferromagnetic state is possible, i.e., $E_0^{N+1}$
corresponds to $S=N/2$ or at least not to $S=0,1/2$. In
Fig. \ref{muplot} we therefore present results only within the regimes
of lowest $S$. In finite systems the onset of Nagaoka state shows a
size dependence and moves to lower values $t^2/U$ with
increasing sizes. From our analysis we can estimate, that the
paramagnetic state with $S=0$ (for even $N+1$) remains stable at least
for $t^2/U > 0.025t$.  Taking mentioned limitations into account, we
can analyse results in Fig. \ref{muplot} which give $\tilde
\mu \sim -0.19t+6.8t^2/U \pm 0.1t$ for $0.025t<t^2/U<0.05t$.
Unfortunately, the proper finite size scaling cannot be performed
since lattice shapes as well as $S$ are not the same for different
$N$ therefore the uncertainty is an estimate. 

Before presenting the final result, we comment on the higher order
corrections to Eq. (\ref{GExpan}) on a triangular lattice.  With the
use of particle-hole transformation we can show that for $\tilde \mu
=0$ there is no correction of order $t^2/U^3$ and the next correction
of order $t^3/U^4$ can be presented as
\begin{equation}
G(\mathbf{k},0)_{\tilde \mu=0}=
  \frac{16}{U^2}
  \varepsilon(\mathbf{k})
   \langle \mathbf{S}_{\delta}\cdot \mathbf{S}_0 \rangle 
   +B(\mathbf k)\frac{t^3}{U^4} .
\end{equation}
We evaluate $B(\mathbf k)$ for some $\mathbf k$ with the use of
numerical calculations on $N=12$ site Hubbard cluster. $B(\mathbf k)$
is obtained by the expansion of numerically calculated GF for the
range $U = 60t - 120t$. Results are presented in Table \ref{tab}. A
relevant observation is that the higher order correction is small in
vicinity of the Luttinger surface for large enough $U>20~t$. Moreover,
the correction is negative and suggests (even) smaller Luttinger
volume, which is the tendency towards larger LSR violation.

\begin{table}
\caption{Numerically obtained expansion coefficients.}
\label{tab}
\begin{center}
\begin{tabular}{ c c }
\hline\noalign{\smallskip}
 k     &  $B(\mathbf k)$\\
\noalign{\smallskip}\hline\noalign{\smallskip}
$(0,0)$ &  $-2320 \pm 40$ \\
$(2 \pi/ 3, 0)$   & $-167 \pm 3$\\
$(\pi , \pi/\sqrt{3})$  & $200\pm5$\\
$(4\pi/3  , 0 )$ & $1810 \pm 40  $\\
\noalign{\smallskip}\hline
\end{tabular}
\end{center}
\end{table}

\begin{figure}
\resizebox{0.5\textwidth}{!}{%
\includegraphics {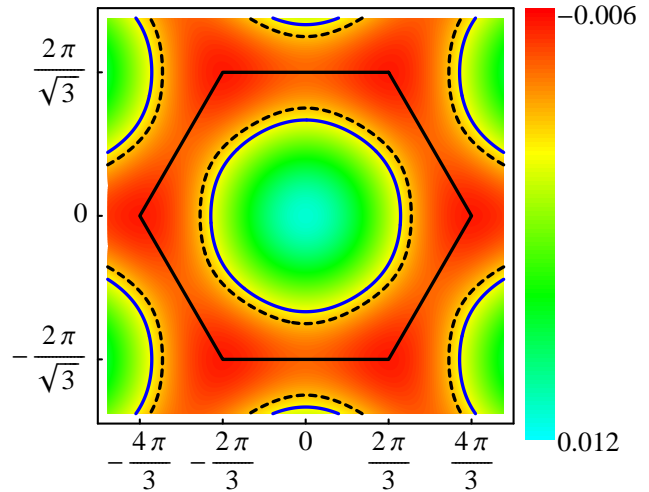}
}
  \caption{(Color online) Green function $G({\bf k},0)$ for 
  $U/t=40$ within 
  the first  Brilloiun zone. Displayed are also the Luttinger volume,
  which would satisfy Luttinger theorem (dashed line) and
  calculated Luttinger volume (full line).}
\label{fig-BC}
\end{figure}

Our final result for GF can be thus presented as
\begin{equation}
G(\mathbf{k},0)= \frac{4}{U^2} \left(4 C_1 \varepsilon(\mathbf{k})
  -0.19t+ 6.8t^2/U \pm 0.1t \right).
\label{GExpan2}
\end{equation}
In Fig. \ref{fig-BC} we show results, Eq. (\ref{GExpan2}), obtained
with the expansion method for particular $U=40t$.  Dashed black line
shows the Luttinger volume, which would correspond to noninteracting
fermions.  The latter can be obtained from Eq. (\ref{GExpan2}) if one
would insert $\tilde \mu_1=-0.61t$ instead of the calculated value.
Evidently, $\tilde \mu_1$ is far from any reasonable estimate in
Fig. \ref{muplot}, where $\tilde \mu(U=40t)\sim 0$. Therefore the
calculated Luttinger volume is obviously smaller than required by the
LSR, the difference being about $17\%$ in Fig. \ref{fig-BC}. Estimated
uncertainty is too small to compensate the violation of the LSR.

\section{Conclusions}
\label{sec:conclu}
In conclusion, we presented evidence that the LSR is violated on
triangular lattice for half-filled Hubbard band in regime of large
$U/t$. Clearly we should stress again limitations to our analysis and
results.  First, $U/t$ must be large enough so that the ground state
corresponds to an insulator with a Mott-Hubbard gap which is the case
for the Hubbard model on triangular lattice at $U \sim 12t$
\cite{aryanpour06}.  On the other side, at very large $U/t$ the onset
of Nagaoka instability appears for the $N+1$ ground state. In absense
of other studies we estimate that this can happen only for $U>40t$.

Finally, let us speculate on the origin of the violation of the LSR.  A
frequent mechanism for the latter is related to the breakdown of the
paramagnetic state or the loss of translational symmetry, in a
microscopic system connected with a phase transition and the onset of
the corresponding long-range order. Besides discussed Nagaoka
instabilities possible at very large $U/t$, within the triangular
lattice another possibility is the long-range antiferromagnetic order
which has been established within the Heisenberg model \cite{bernu92}
but so far not within the Hubbard model at half filling
\cite{bulut05}.  The onset of such order would limit the
relevant $U/t$ parameter window of the paramagnetic
insulating state. On the other hand, any breaking of translational
symmetry is expected to change entirely the Fermi surface topology and
not to partially reduce its volume as found in our study.  

More radical conclusion would be that the LSR is generally violated
within the Mott-Hubbard insulators without the particle-hole symmetry
\cite{stanescu07,rosch07}, and in particular within the insulating
state of the Hubbard model on a triangular lattice. We note that the
main formal requirement for the validity of the LSR
\cite{luttinger60b} is the adiabatic development of the ground state
(i.e. of the free energy in the limit $T\to 0$, where the LSR becomes
applicable) 
with the increasing electron-electron repulsion. Hubbard model
is perturbative in $U$ and adiabatically connected to the
noninteracting fermion system at least for weak $U/t$. On the other
hand, the metal-insulator transition at $U=U_c$ seems to 
represent a point, where the perturbation theory as well as LSR breaks
down. Such scenario is still far from obvious so further studies in
this direction are needed.


\end{document}